# Beam-Based Determination of the Offset of Booster $\gamma_T$ Quads

Xi Yang, Charles M. Ankenbrandt, and James MacLachlan

*Fermi National Accelerator Laboratory*

Box 500, Batavia IL 60510

## Abstract

Twelve pulsed $\gamma_T$ quads have been installed in the Booster to provide fast transition crossing. The less time the beam stays in the non-adiabatic period near transition, the less the longitudinal emittance grows. From the past experience, the $\gamma_T$ quads are not well aligned relative to the usual closed orbit. Quad steering can cause beam loss and a dispersion wave after transition. To make the $\gamma_T$ quads routinely operational, procedures for finding the center of the beam relative to the quads and centering the beam through all of them are very important. A program, which uses the difference in the closed orbits when $\gamma_T$ quads are on and off and calculates the offsets of the beam relative to $\gamma_T$ quads, has been developed and tested. A radial orbit offset (ROF) of about 3 mm has been experimentally determined to be nearly the optimal radial position for centering the beam through all the $\gamma_T$ quads, thereby eliminating the immediate need for repositioning the quads.

## Introduction

It has been very helpful to make a specialized program to analyze the quad steering at known positions in the Fermilab Booster from the difference orbit taken with the $\gamma_T$ quads on and off. There are 48 beam position monitors (BPM) used in Booster for monitoring the beam positions both horizontally and vertically.[1] Twelve pulsed $\gamma_T$ quads, which were designed to change $\gamma_T$ one unit within 100 μs, have been installed to make the transition crossing faster to avoid longitudinal emittance growth. Because they are not perfectly aligned, they steer the beam and can cause beam loss when they are pulsed. A program has been developed and tested for finding the offsets of $\gamma_T$ quads by calculating the angle kicks at $\gamma_T$ quad locations from the difference orbit taken from the BPM data. Once angle kicks at $\gamma_T$ quad locations are known from the program, beam offsets at $\gamma_T$ quad locations can be calculated with known $\gamma_T$ quad parameters. Either by adding position bumps or by moving the quads, the $\gamma_T$ quad steering can be minimized.

Because we want to find twelve $\gamma_T$ quad angle kicks from forty-eight BPM readings, the least squares method has been used in the analysis. The calculated quad kicks are used to predict the difference orbit, and the result is compared to the BPM data. The predicted difference orbit matches that measured from the BPM data quite well, and the root-mean-square (RMS) is less than 0.2 mm at the ROF value of 3.

### Method

The transverse displacement ($\Delta x$) introduced by a dipole kick ($\theta$) is

$$\Delta x(s_i) = \sum_{j=1}^{n}\left(\sqrt{\beta(s_j)\times\beta(s_i)}\times\theta(s_j)\times\cos\left(\pi\nu_x-\left|\mu(s_i)-\mu(s_j)\right|\right)\Big/\left(2\times\sin(\pi\times\nu_x)\right)\right)=A_{ij}\times\theta(s_j)$$

(1)

Here, $s_i$ is the longitudinal position (LP) where the transverse displacement is observed, and $s_j$ is the LP of the angle kick. $\left|\mu(s_i)-\mu(s_j)\right|$ is the phase advance between $s_i$ and $s_j$. $\nu_x$ is the betatron tune in the $x$ direction. $n$ is the total number of angle kicks. When the number of places where transverse displacements are observed is greater than the number of places where angle kicks appear, there is more than enough information to solve eq.(1) for the angular kicks. The least squares method is used to find the optimal solution. Eq.(2) is used to find the angle kicks from the measured difference orbit.

$$\theta_k = \sum_{j=1}^{n}\left(\Re_{jk}\right)^{-1}B_j$$

(2)

$n = 12$ is the number of angle kicks; the matrix $\Re$ is

$$\Re_{jk} = \sum_{i=1}^{m}A_{ij}A_{ik}\ .$$

$m = 48$ is the number of BPM's; $A$ is a 48 by 12 matrix defined in eq. 1. Array $B$ is

$$B_j = \sum_{i=1}^{m}\Delta x_i A_{ij}\ .$$

$\Delta x_i$ is the difference orbit between conditions of $\gamma_T$ quads on and off at LP $s_i$.

$x_j$ is the offset of the center of the beam relative to the center of the $\gamma_T$ quad at LP $s_j$:

$$x_j = \theta_j\Big/(k_j\times l_j)$$

(3)

$l_j$ is the length of the $\gamma_T$ quad, and $k_j$ is the quad strength. $\beta(s_i)$, $\beta(s_j)$, $\mu(s_i)$ $\mu(s_j)$, and $\nu_x$ are obtained from the Booster lattice file, which is calculated using MAD.[2]. Also, $k_j$ and $l_j$ are obtained from the MAD input file.

**Experimental Results**

All the experiments were done at the extracted beam intensity of $0.315 \times 10^{12}$ protons. A programmed radial offset (ROF) was used to move the beam radially when the $\gamma_T$ quads were pulsed to establish the ROF setting where the beam was best centered through all the $\gamma_T$ quads. The difference orbit between $\gamma_T$ quads on and off was found at eight different ROF values, -2, -1, 0, 1, 2, 3, 4, 5; the results are shown in Fig. 1. All the orbits with $\gamma_T$ quads on were taken at the same $\gamma_T$ quad current of 780 A. The black, red, green, blue, cyan, magenta, yellow, dark yellow curves represent eight different ROF values of -2, -1, 0, 1, 2, 3, 4, 5 respectively. It is clear that the beam is best centered through all the $\gamma_T$ quads at the ROF value of 3. The angle kicks at the 12 $\gamma_T$ quad locations were calculated from the difference orbits in Fig. 1 using the least squares program, and the results were used for predicting the difference orbit. The black, red curves in Figs. 2(a)-2(h) represent the predicted difference orbit and the measured difference orbit from BPM data at eight different ROF values of -2 to 5 respectively. The differences between the measured difference orbits and the predicted difference orbits at eight ROF values of -2 to 5 are shown in Fig. 3(a). The predicted and measured difference orbits agree with each other quite well, especially at the ROF value of 3, and the RMS of their differences are shown in Fig. 3(b). The calculated angle kicks at 12 $\gamma_T$ quad locations with ROF values of -2 to 5 are shown Fig. 4(a), and the calculated offsets of the beam relative to $\gamma_T$ quads are shown in Fig. 4(b). The difference orbits, angle kicks, and offsets in the vertical direction are shown in Fig. 5(a)-5(c) respectively.

**Conclusion**

The angle kicks from twelve pulsed $\gamma_T$ quads have been calculated from the difference orbit taken with $\gamma_T$ quads on and off using a least squares fitting technique. The ROF value of 3 (about 3 mm) has been experimentally determined to be the optimal radial offset for centering the beam through all the $\gamma_T$ quads at the time they are pulsed.

**Acknowledgment**

Thanks to Dr. Alexandr Drozhdin for providing the Booster lattice file and performing MAD calculations of lattice functions at the particular time and parameters relevant to our difference orbits.

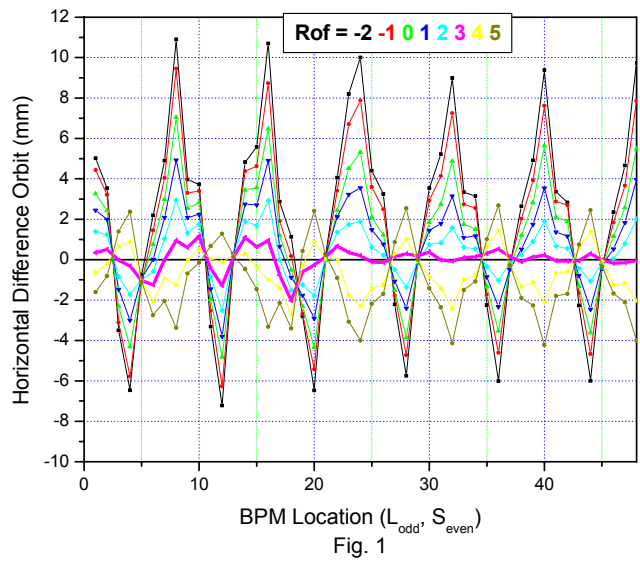

Fig. 1 The difference orbit in the horizontal direction with $\gamma_T$ quads on and off measured at eight ROF values for the extracted beam intensity of $0.315 \times 10^{12}$ protons. The black, red, green, blue, cyan, magenta, yellow, dark yellow curves represent the eight different ROF values of -2, -1, 0, 1, 2, 3, 4, 5 respectively.

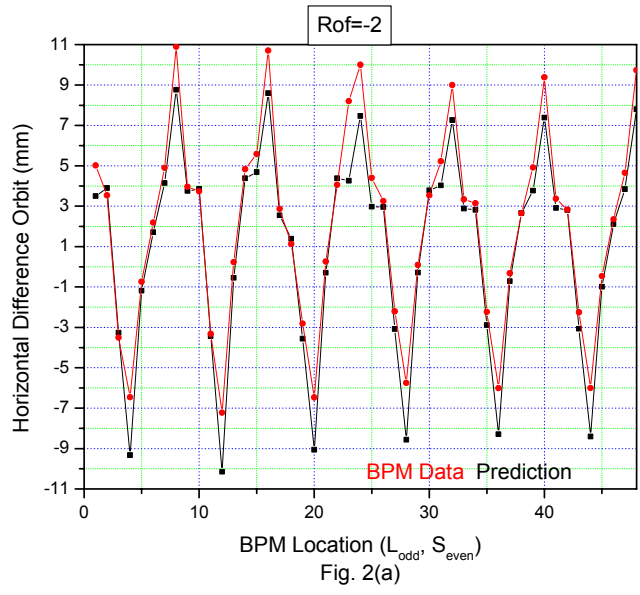

Fig. 2(a)

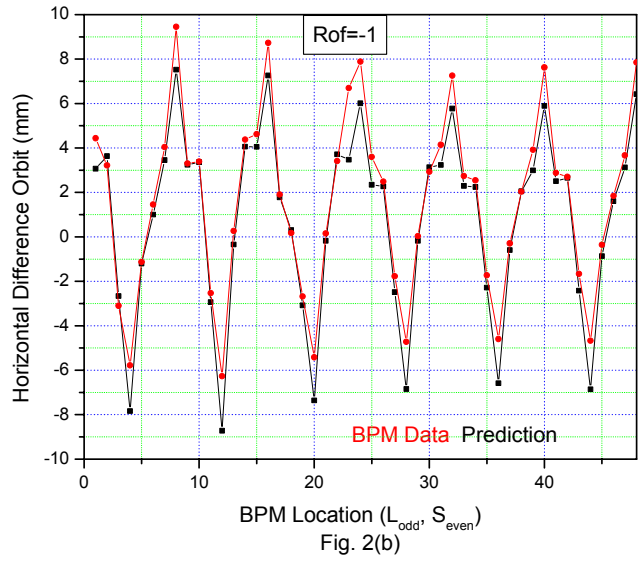

Fig. 2(b)

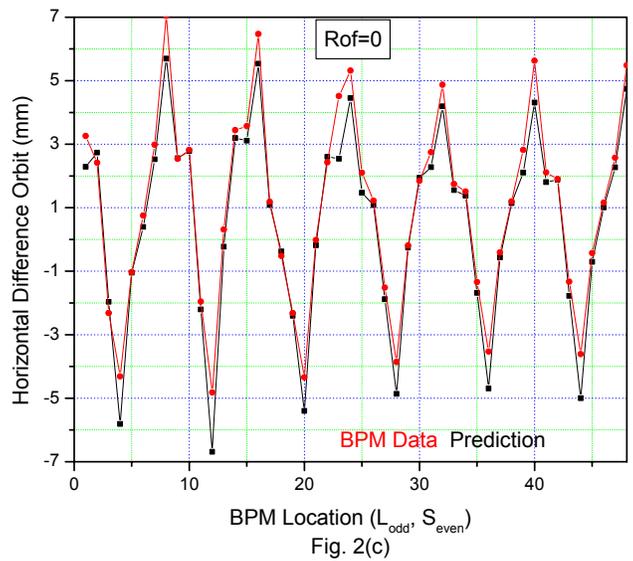

Fig. 2(c)

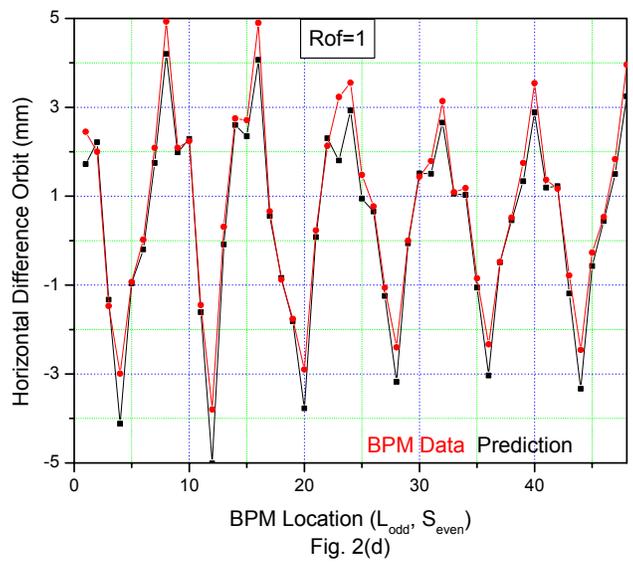

Fig. 2(d)

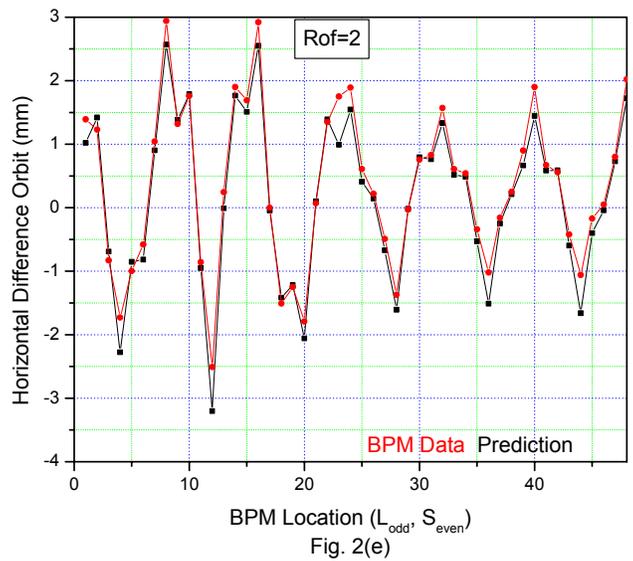

Fig. 2(e)

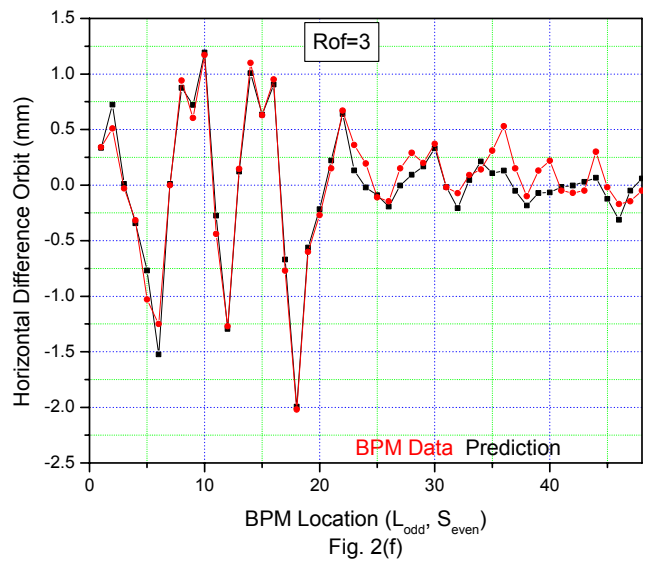

Fig. 2(f)

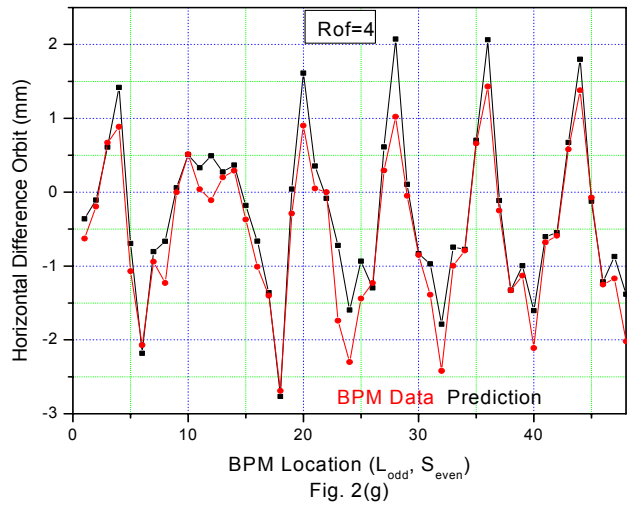

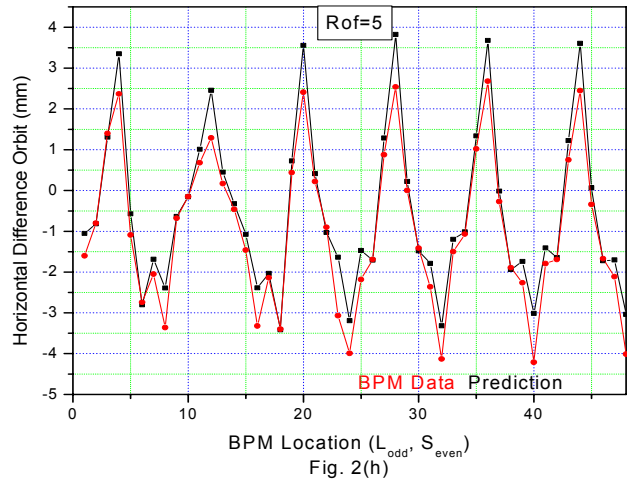

Fig. 2(a) the black and red curves represent the predicted difference orbit and the measured difference orbit using horizontal BPM readings at ROF of -2.

Fig. 2(b) ROF of -1.

Fig. 2(c) ROF of 0.

Fig. 2(d) ROF of 1.

Fig. 2(e) ROF of 2.

Fig. 2(f) ROF of 3.

Fig. 2(g) ROF of 4.

Fig. 2(h) ROF of 5.

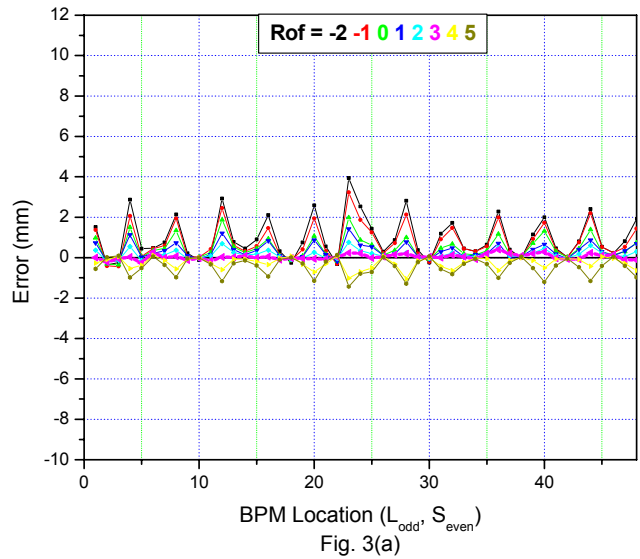

Fig. 3(a)

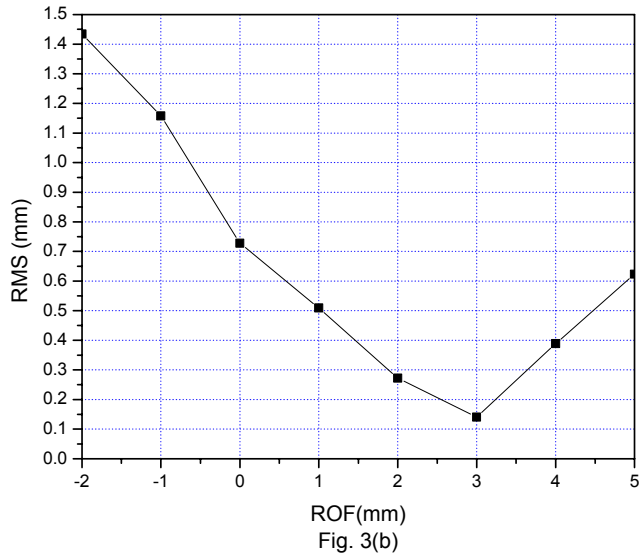

Fig. 3(b)

Fig. 3(a) the difference between the measured difference orbit and the predicted difference orbit in the horizontal direction at eight different ROF values of -2 to 5. The black, red, green, blue, cyan, magenta, yellow, dark yellow curves represent the eight different ROF values of -2, -1, 0, 1, 2, 3, 4, 5 respectively.

Fig. 3(b) the RMS of the difference between the predicted difference orbit and the measured difference orbit at eight different ROF values.

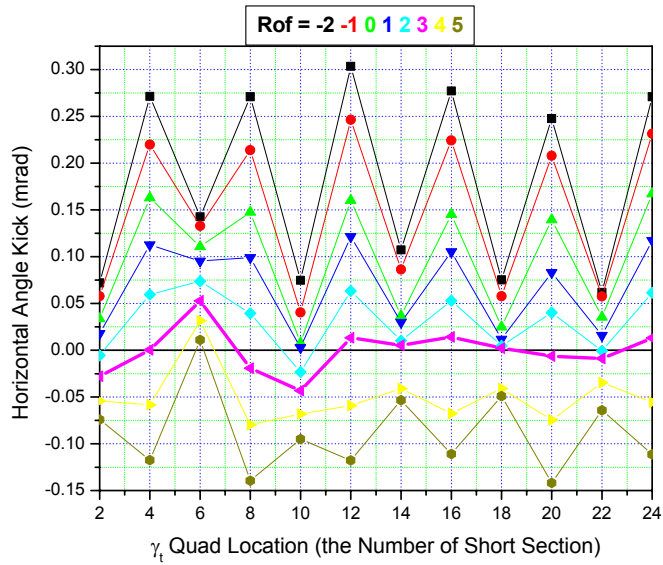

Fig. 4(a)

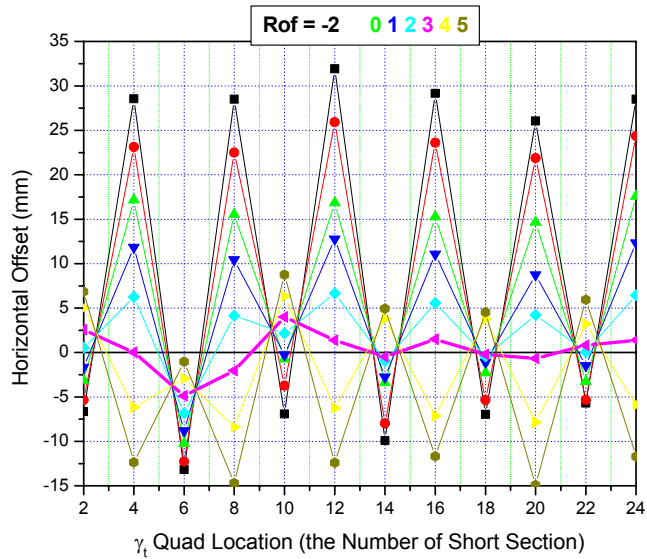

Fig. 4(b)

Fig. 4(a) the calculated angle kicks in the horizontal direction at 12 $\gamma_T$ quad locations with ROF values of -2 to 5. The black, red, green, blue, cyan, magenta, yellow, dark yellow curves represent the eight different ROF values of -2, -1, 0, 1, 2, 3, 4, 5 respectively.

Fig. 4(b) the calculated offsets of the beam relative to $\gamma_T$ quads in the horizontal direction at ROF values of -2 to 5.

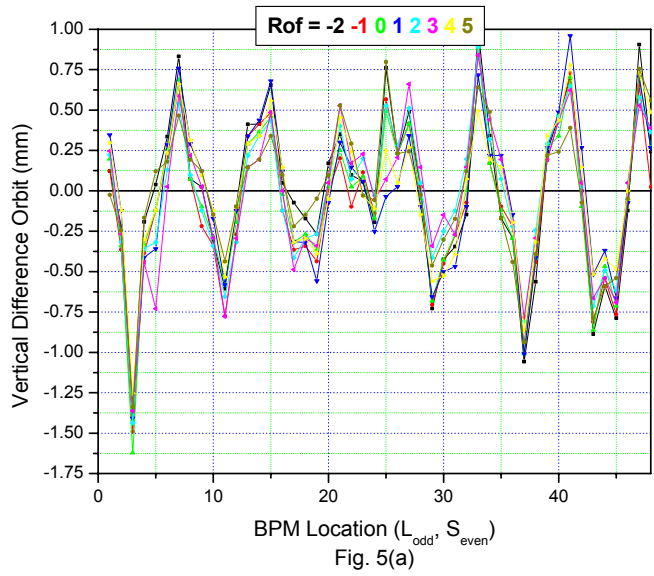

Fig. 5(a)

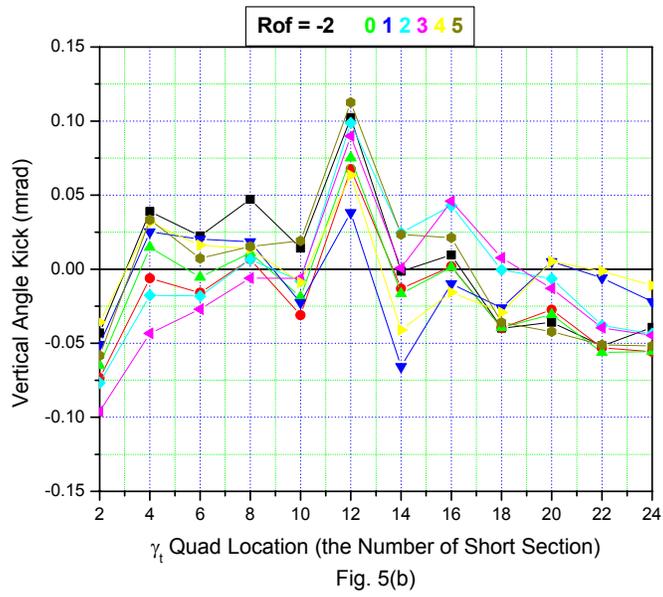

Fig. 5(b)

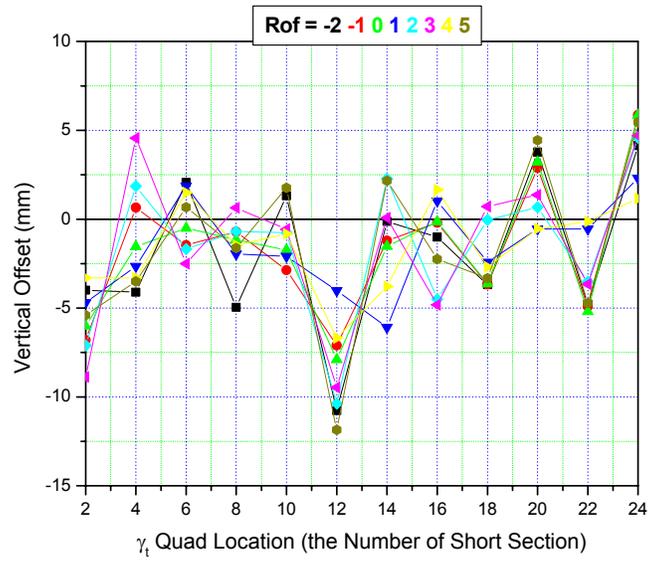

Fig. 5(c)

Fig. 5(a) the difference orbits in the vertical direction at eight ROF values of -2 to 5. The black, red, green, blue, cyan, magenta, yellow, dark yellow curves represent the eight different ROF values of -2, -1, 0, 1, 2, 3, 4, 5 respectively.

Fig. 5(b) the angle kicks in the vertical direction at eight ROF values of -2 to 5.

Fig. 5(c) the calculated offsets of the beam relative to $\gamma_T$ quads in the vertical direction at ROF values of -2 to 5.